\documentclass[aps,prl,twocolumn,showpacs,a4paper]{revtex4}

\usepackage{amsfonts}
\usepackage{amsmath}
\usepackage{dcolumn}
\usepackage[english]{babel}
\usepackage[pdftex]{graphicx}

\begin{document}
\title{Nonlinear elasticity of monolayer graphene}
\author{Emiliano Cadelano, Pier Luca Palla, Stefano Giordano, Luciano Colombo}
\email[e-mail: ]{luciano.colombo@dsf.unica.it}
\affiliation{Dipartimento di Fisica, Universit\`a di Cagliari and\\ SLACS-INFM/CNR Sardinian Laboratory for Computational Materials Science \\ Cittadella Universitaria, I-09042 Monserrato (Ca), Italy}
\date{\today}

\begin{abstract}
By combining continuum elasticity theory and tight-binding atomistic simulations, we work out the constitutive nonlinear stress-strain relation for graphene stretching elasticity and we calculate all the corresponding nonlinear elastic moduli. Present results represent a robust picture on elastic behavior of one-atom thick carbon sheets and provide the proper interpretation of recent experiments. In particular, we discuss the physical meaning of the effective nonlinear elastic modulus there introduced and we predict its value in good agreement with available data. Finally, a hyperelastic softening behavior is observed and discussed, so determining the failure properties of graphene.
\end{abstract}
\maketitle

\begin{figure}[b]
\includegraphics[scale=0.32]{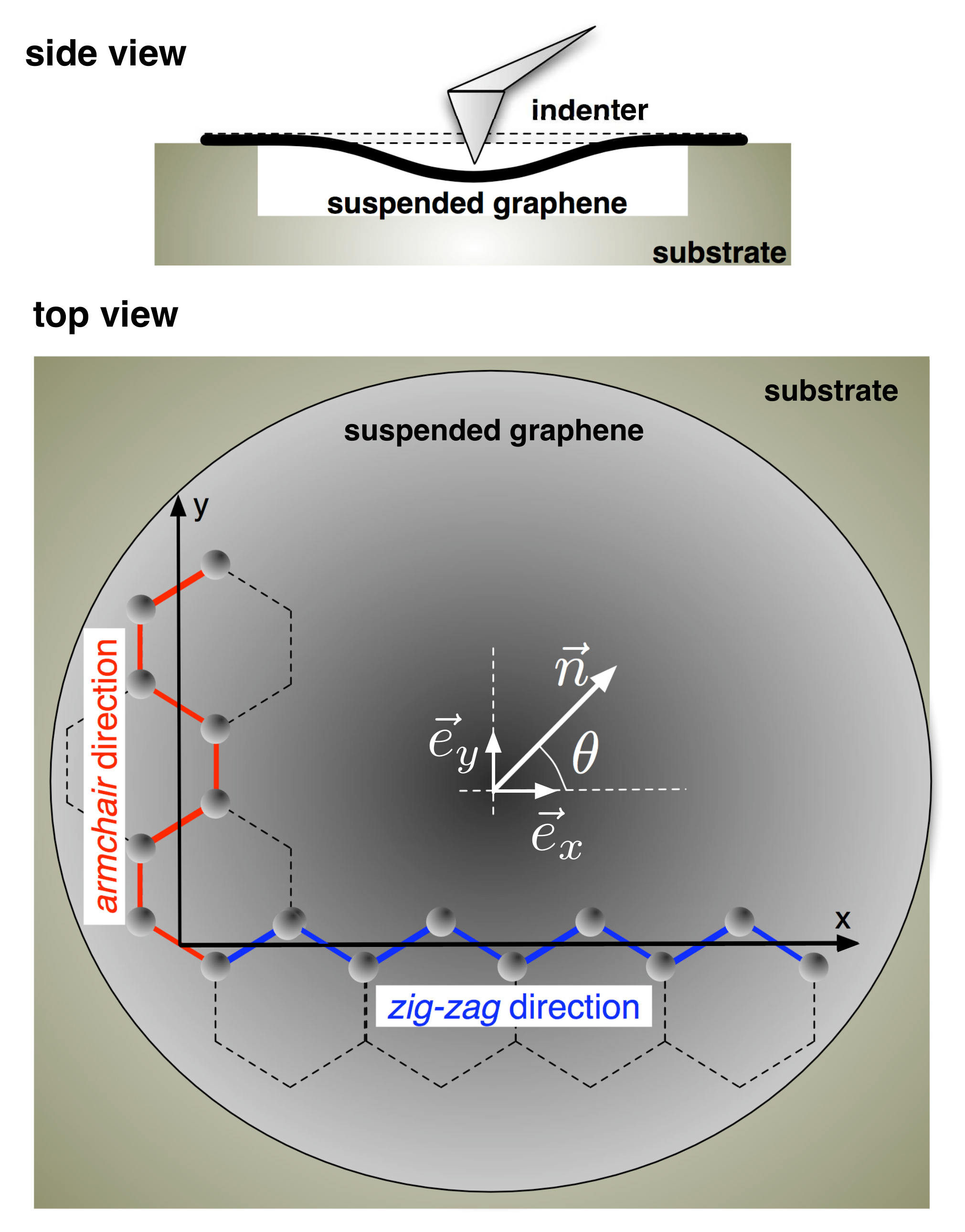}
\caption{
\label{expt} (color online)
Top: schematic representation of the indentation of a suspended monolayer graphene (side view). Bottom: definition of zig-zag and armchair directions (top view). The gray-scale shading of the monolayer graphene pictorially represents the radially symmetric strain field generated by indentation.}
\end{figure}

The elastic properties of graphene have been recently determined by atomic force microscope nanoindentation \cite{lee,gomez}, measuring the deformation of a free-standing monolayer  as sketched in Fig.\ref{expt} (top). In particular, in Ref.\cite{lee} the experimental force-deformation relation has been expressed as a phenomenological nonlinear scalar relation between the applied stress ($\sigma$) and the observed strain ($\epsilon$)
\begin{equation}
\sigma = E \epsilon + D \epsilon^2
\label{av}
\end{equation}
where $E$ and $D$ are, respectively, the Young modulus and an effective nonlinear (third-order) elastic modulus of the two dimensional carbon sheet. The reported experimental values are: $E=340\pm40$ Nm$^{-1}$ and $D=-690\pm120$Nm$^{-1}$. While the first result is consistent with previous existing data \cite{zhou,arroyo,michel,kudin,tu}, the above value for $D$ represents so far the  only available information about the nonlinear elasticity of a one-atom thick carbon sheet. 

Although nonlinear features are summarized in Eq.(\ref{av}) by one effective parameter $ D $, continuum elasticity theory predicts the existence of three independent third-order parameters $ C_{ijk} $ for graphene, as reported below. In other words, while Eq.(\ref{av}) represents a valuable effective relation for the interpretation of a complex experiment \cite{lee}, it must be worked out a more rigorous theoretical picture in order to properly define all the nonlinear elastic constants of graphene and to understand the physical meaning of $D$. This corresponds to the content of the present Letter where we investigate the constitutive nonlinear stress-strain relation of graphene, by combining continuum elasticity and tight-binding atomistic simulation (TB-AS) \cite{colombo}. 

To obtain the nonlinear stress-strain relation of an elastic membrane, we need at first to elaborate an expression for the corresponding strain energy function $U$ (per unit area). Since, as illustrated in Fig.1(bottom), the underlying lattice is hexagonal, it is useful to consider the coordinate set $ \alpha = x + iy $ and $ \beta = x - iy $ \cite{landau}, where the $x$ and $y$ directions are respectively identified with the zig-zag ({\it zz}) and the armchair ({\it ac}) directions. Because the strain energy function is invariant under a rotation of $\pi/3 $ about
the $z$-axis (normal to the suspended monolayer), there are two linear moduli (the two-dimensional Young modulus $E$ and Poisson ratio $ \nu $) and three nonlinear independent elastic coefficients ($ \Lambda_i $, $i=1,2,3$) all expressed in units of force/length; we easily proved that
\begin{eqnarray}
\nonumber
2U&=& \frac{E}{1+\nu} \epsilon_{\alpha \alpha}\epsilon_{\beta \beta}+\frac{E\nu}{1-\nu^{2}} \epsilon^{2}_{\alpha \beta}\\
\label{strainenergy}
&+&\Lambda_1(\epsilon^{3}_{\alpha \alpha}+\epsilon^{3}_{\beta \beta}) +\Lambda_2 \epsilon_{\alpha \alpha}\epsilon_{\beta \beta}\epsilon_{\alpha \beta} +\Lambda_3 \epsilon^{3}_{\alpha \beta},
\end{eqnarray}
where $\epsilon_{\alpha \beta}= \epsilon_{xx}+\epsilon_{yy} $, $ \epsilon_{\alpha \alpha}= \epsilon_{xx}-\epsilon_{yy} +2i\epsilon_{xy}$, and $ \epsilon_{\beta \beta}= \epsilon_{xx}-\epsilon_{yy} -2i\epsilon_{xy} $. In order to further proceed we must better focus the strain definition which in elasticity theory is twofold: we can introduce the so-called small strain tensor $\hat{\epsilon}=\frac{1}{2}(\vec{\nabla}\vec{u}+\vec{\nabla}\vec{u}^{\rm T})$, being $\vec{u}$ the displacement field, or the Lagrangian strain $\hat{\eta}=\frac{1}{2}(\vec{\nabla}\vec{u}+\vec{\nabla}\vec{u}^{\rm T}+\vec{\nabla}\vec{u}^{\rm T}\vec{\nabla}\vec{u})$. While $\hat{\epsilon}$ takes into account only the physical nonlinearity features (it describes a nonlinear stress-strain dependence observed in regime of small deformation), $\hat{\eta}$ describes any possible source of nonlinearity, i.e. it includes both physical and geometrical (large deformation) ones.

We start using $\hat{\epsilon}$ in Eq.(\ref{strainenergy}) and we get the nonlinear elastic coefficients $ \Lambda_1 $, $ \Lambda_2 $ and $ \Lambda_3 $ which are related to the third order elastic constants $ C_{111} $, $ C_{222} $ and $ C_{112} $, as customarily defined in crystal elasticity \cite{huntington}, through the following relations
\begin{eqnarray}
\label{lambda}
\nonumber
\Lambda_{1} =\frac{1}{12}(C_{111}-C_{222}),\,\,\,\,\,\,\,\,\,\,
\Lambda_{2} =\frac{1}{4}(C_{222}-C_{112}),\\
\Lambda_{3} =\frac{1}{12}(2C_{111}-C_{222}+3C_{112}).\,\,\,\,\,\,\,\,\,\,\,\,\,\,\,\,\,
\end{eqnarray}
The strain energy function is finally obtained as 
\begin{eqnarray}
\nonumber
2U&=& \frac{E}{1+\nu} \mbox{Tr}\left( \hat{\epsilon}^{2}\right) +\frac{E\nu}{1-\nu^{2}} \left( \mbox{Tr} \hat{\epsilon}\right)^{2}
\\ 
\nonumber
&&+\frac{1}{3} C_{111}\epsilon_{xx}^{3}+\frac{1}{3} C_{222}\epsilon_{yy}^{3} +C_{112} \epsilon_{xx}^{2} \epsilon_{yy}\\
\nonumber
&&+(C_{111}-C_{222}+C_{112})\epsilon_{xx} \epsilon_{yy}^{2}\\
\nonumber
&&+(3C_{222}-2C_{111}-C_{112})\epsilon_{xx} \epsilon_{xy}^{2}\\
\label{complete}
&&+(2C_{111}-C_{222}-C_{112})\epsilon_{yy} \epsilon_{xy}^{2}
\end{eqnarray}
where we set $ \epsilon_{\alpha \alpha}\epsilon_{\beta \beta}=\mbox{Tr}\left( \hat{\epsilon}^{2}\right)  $ and $ \epsilon^{2}_{\alpha \beta}=\left( \mbox{Tr} \hat{\epsilon}\right)^{2} $.
The stress-strain nonlinear constitutive equation for in-plane stretching is straightforwardly obtained by $
\hat{T}= \partial U/ \partial \hat{\epsilon}$, where $ \hat{T} $ is the Cauchy stress tensor. 

Since the analysis of the experimental data provided in Ref.\cite{lee} through Eq.(\ref{av}) is assuming an applied uniaxial stress, we now suppose to apply a uniaxial tension 
$ \sigma_{\vec{n}} $ along the arbitrary direction $ \vec{n}=\cos\theta \vec{e}_{x}+\sin\theta \vec{e}_{y} $, where $ \vec{e}_{x} $ and $ \vec{e}_{y} $ are the unit vectors along the zig-zag and the armchair directions, respectively (see Fig.\ref{expt} , bottom). Under this assumption we get:
$ \hat{T}=\sigma_{\vec{n}}  \vec{n}\otimes\vec{n}$, with in-plane components defined as $ T_{xx}=\sigma_{\vec{n}} \cos^{2}\theta $, $T_{xy}=\sigma_{\vec{n}}\cos\theta\sin\theta$, and $ T_{yy}=\sigma_{\vec{n}} \sin^{2}\theta $. Similarly, by inverting the nonlinear constitutive equation we find the corresponding strain tensor and the relative variation of length $\epsilon_{\vec{n}}= \vec{n}\cdot \hat{\epsilon} \vec{n} $ along the direction $ \vec{n} $. By combining these results, we obtain the stress-strain relation along the arbitrary direction ${\vec{n}}$ 
\begin{eqnarray}
\sigma_{\vec{n}}=E\epsilon_{\vec{n}}+D_{\vec{n}}\epsilon_{\vec{n}}^{2}
\label{constn}
\end{eqnarray}
where $ D_{\vec{n}} $ is given by
\begin{eqnarray}
\label{dn}
D_{\vec{n}}=\frac{3}{2}\left(1-\nu\right)^{3}\Lambda_{3}+\frac{3}{2}\left(1-\nu\right)\left(1+\nu\right)^{2}\Lambda_{2} \,\,\,\,\,\,\,\,\,\,\,\,\,\,\\
+3\left( 2\cos^{2}\theta-1\right)\left(16\cos^{4}\theta-16\cos^{2}\theta+1\right) \left(1+\nu\right)^{3}\Lambda_{1}
\nonumber
\end{eqnarray}
If we set  $\vec{n}=\vec{e}_{x}  $ (i.e. $ \theta=0 $), we get the nonlinear modulus 
$D^{(zz)} $ for stretching along the zig-zag direction
\begin{eqnarray}
\nonumber
D^{(zz)}=D_{\vec{e}_{x}}&=&3\left(1+\nu\right)^{3}\Lambda_{1}+\frac{3}{2}\left(1-\nu\right)\left(1+\nu\right)^{2}\Lambda_{2} \\
\label{dz}
&&+\frac{3}{2}\left(1-\nu\right)^{3}\Lambda_{3}
\end{eqnarray}
Similarly, by setting $ \vec{n}=\vec{e}_{y}  $ (i.e. $ \theta=\pi/2 $), we obtain the nonlinear modulus $D^{(ac)}$ for stretching along the armchair  direction
\begin{eqnarray}
\nonumber
D^{(ac)}=D_{\vec{e}_{y}}&=&-3\left(1+\nu\right)^{3}\Lambda_{1}+\frac{3}{2}\left(1-\nu\right)\left(1+\nu\right)^{2}\Lambda_{2} \\
\label{da}
&&+\frac{3}{2}\left(1-\nu\right)^{3}\Lambda_{3}
\end{eqnarray}
We observe that the above expression for $D^{(zz)}$ apply for all stretching directions defined by the angles $\theta=k\pi/3$ ($k \in \mathbb{Z}$), while  $D^{(ac)}$ holds for the angles $\theta=\pi/6+k\pi/3$.

Since the nanoindentation experiments generate a strain field with radial symmetry \cite{lee}, as sketched in Fig.\ref{expt}(bottom), in order to get the unique {\it scalar} nonlinear elastic modulus appearing in Eq.(1) we need to average the expression of $ D_{\vec{n}} $ over $\theta $. This procedure leads to 
\begin{eqnarray}
\nonumber
\langle D_{\vec{n}}\rangle&=&\frac{1}{2\pi}\int_{0}^{2\pi}D_{\vec{n}} {\rm d} \theta=\frac{D^{(zz)}+D^{(ac)}}{2}\\
&=&\frac{3}{2}\left(1-\nu\right)\left[ \left(1+\nu\right)^{2}\Lambda_{2} +\left(1-\nu\right)^{2}\Lambda_{3}\right] 
\label{dav}
\end{eqnarray}
proving that the experimentally determined nonlinear modulus actually corresponds to the average value of the moduli for the zig-zag and armchair directions. 

We now repeat the above procedure by using the Lagrangian strain $\hat{\eta}$: even in this case it is demonstrated that the strain energy function is given by the very same Eq.(\ref{complete}) where $ \hat{\epsilon} $ is replaced by $ \hat{\eta} $ and the $ C_{ijk} $ by the Lagrangian third-order moduli $ C^{\mathcal{L}}_{ijk} $. By imposing the identity $ U(\hat{\epsilon})=U(\hat{\eta}) $ (where the Lagrangian strain can be written in term of the small strain by $ \hat{\eta}=\hat{\epsilon}+\frac{1}{2}\hat{\epsilon}^{2} $ \cite{lag1,lag2}) we obtain the conversion rules: $ C^{\mathcal{L}}_{111}=C_{111}-\frac{3E}{1-\nu^{2}} $, $ C^{\mathcal{L}}_{222}=C_{222}-\frac{3E}{1-\nu^{2}} $, $ C^{\mathcal{L}}_{112}=C_{112}-\frac{E\nu}{1-\nu^{2}} $, $ D^{\mathcal{L}}_{\vec{n}}= D_{\vec{n}}-\frac{3}{2}E$ (for any $ \vec{n} $) and $\langle D^{\mathcal{L}}_{\vec{n}} \rangle=\langle D_{\vec{n}} \rangle -\frac{3}{2}E$. The constitutive equation can be finally derived in the form $\hat{T}^{\mathcal{PK}}= \partial U/ \partial \hat{\eta}$, where $ \hat{T}^{\mathcal{PK}} $ is the second Piola-Kirchhoff stress tensor. Hereafter we will refer to the small strain and Lagrangian scalar nonlinear modulus by  $\langle D_{\vec{n}} \rangle$ and $\langle D^{\mathcal{L}}_{\vec{n}} \rangle$, respectively. They both will be compared with the experimental parameter $ D $ of Eq.(\ref{av}). The analysis below will identify the actual theoretical counterpart of $ D $.

The important result summarized in Eq.(\ref{dav}) (as well as in its Lagrangian version) implies that the scalar nonlinear modulus can be obtained by the third-order elastic constants (as well as the linear ones). They can be computed through the energy-vs-strain curves corresponding to suitable homogeneous in-plane deformations, thus avoiding a technically complicated simulation of the nanoindentation experiment. Therefore, the following in-plane deformation have been applied: (i) an uniaxial deformation $\zeta$ along the zig-zag direction, corresponding to a strain tensor
$\epsilon_{ij}^{(zz)}=\zeta\delta_{ix}\delta_{jx}$; (ii) an uniaxial deformation $\zeta$ along the armchair direction, corresponding to a strain tensor $\epsilon_{ij}^{(ac)}=\zeta\delta_{iy}\delta_{jy}$; (iii) an hydrostatic planar deformation $\zeta$, corresponding to the strain tensor $\epsilon_{ij}^{(p)}=\zeta\delta_{ij}$; (iv)  a shear deformation $\zeta$, corresponding to an in-plain strain tensor $\epsilon_{ij}^{(s)}=\zeta \left(\delta_{ix}\delta_{jy}+\delta_{iy}\delta_{jx} \right) $.

\begin{table}[t]
\caption{\label{table1} Relationship among the energy expansion coefficients $U^{(2)}$ and $U^{(3)}$ of Eq.(\ref{fit}) and the elastic moduli of graphene for four in-plane deformations (see text).}
\begin{tabular}{ccccc}
\hline
\hline	
deformation	   &	\ \ \ \ \ & $U^{(2)}$ 		& \ \ \ & $U^{(3)}$ \\
\hline
$\epsilon_{ij}^{(zz)}$  \ \ \ \ \ & & $\frac{E}{1-\nu^{2}}$ &\ \ \ \ &	$C_{111}$	\\
$\epsilon_{ij}^{(ac)}$ \ \ \ \ \ & & $\frac{E}{1-\nu^{2}}$ &\ \ \ \  &	$C_{222}$	\\
$\epsilon_{ij}^{(p)}$ \ \ \ \ \ & & $\frac{2E}{1-\nu}$	& \ \ \ \  &	$4C_{111}-2C_{222}+6C_{112}$	\\
$\epsilon_{ij}^{(s)}$ \ \ \ \ \ & & $\frac{2E}{1+\nu}$	 & \ \ \ \ &	$0$	\\
\hline
\hline
\end{tabular}
\end{table}

\begin{figure}[b]
 \begin{center}
\begin{tabular}{c}
\resizebox{88mm}{!}{\includegraphics{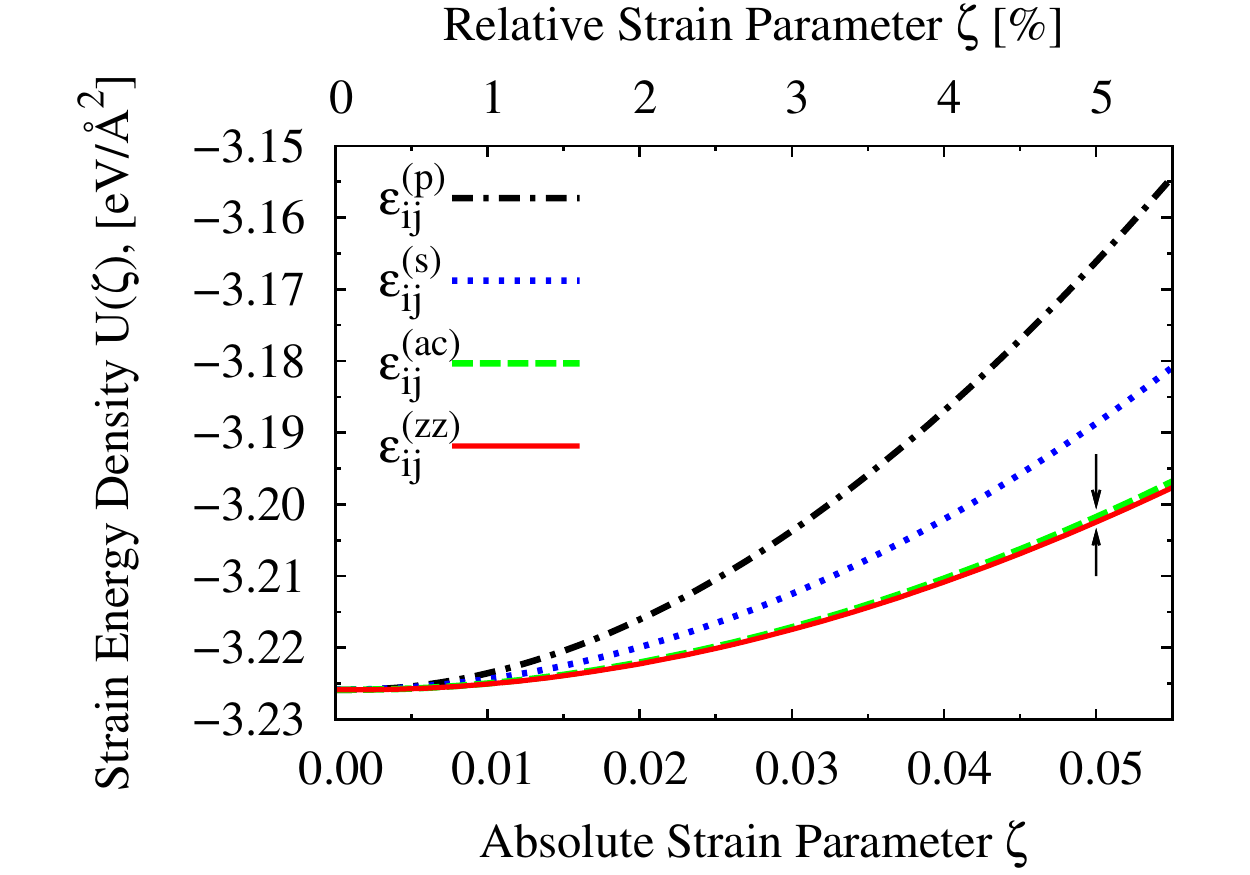}}
\end{tabular}
\caption{\label{energy} (color online) Strain energy density $ U $, obtained by TB-AS, as function of the strain parameter $ \zeta $ corresponding to the four homogeneous deformations summarized in Tab.\ref{table1}.}
\end{center}
\end{figure}

In this work the needed energy-vs-strain curves have been determined by TB-AS, making use of the tight-binding representation by Xu et al. \cite{xu}. A periodically repeated square cell containing 400 carbon atoms was deformed as above. For any given applied deformation, full relaxation of the internal degrees of freedom of the simulation cell was performed by zero temperature damped dynamics  until interatomic forces resulted not larger than $0.5\cdot 10^{-11}$eV/\AA. Similar calculations were repeated by using a smaller simulation cell containing 200 atoms and by relaxing the system through simulated annealing. No deviation from data here reported were observed.

\begin{table}[t]
\caption{\label{table2} Small strain and Lagrangian nonlinear elastic moduli of graphene in units of Nm$^{-1}$. }
\begin{tabular}{cdccd}
\hline
 \hline
 	\multicolumn{2}{c}{Small strain} & 	 &  \multicolumn{2}{c}{Lagrangian} 	 \\
 \hline
    $C_{111}$	& -1689.2		&$ \,\,\,\,\,\,\, $& $C^{\mathcal{L}}_{111}$				&	-2724.7	\\
    $C_{222}$	& -1487.7		&$ \,\,\,\,\,\,\, $& $C^{\mathcal{L}}_{222}$		     	&	-2523.2  \\
    $C_{112}$	& -484.1		    &$ \,\,\,\,\,\,\, $& $C^{\mathcal{L}}_{112}$		     	&	-591.1	\\
    $D^{(zz)}$	& -696.2  		&$ \,\,\,\,\,\,\, $& $D^{\mathcal{L}(zz)}$		    	    &	-1163.7	\\
    $D^{(ac)}$	& -469.6			&$ \,\,\,\,\,\,\, $& $D^{\mathcal{L}(ac)}$			        &	-937.9\\
\hline
\hline
\end{tabular}
\end{table}

For the deformations $\epsilon_{ij}^{(zz)}$, $\epsilon_{ij}^{(ac)}$, $\epsilon_{ij}^{(p)}$ and $\epsilon_{ij}^{(s)}$  the elastic energy of strained graphene can be written in terms of just the single deformation parameter 
$\zeta$
\begin{equation}
 \label{fit}	
U(\zeta)=U_0+\frac{1}{2}U^{(2)}\zeta^2+\frac{1}{6}U^{(3)}\zeta^3+O(\zeta^4)
\end{equation}
where $U_0$ is the energy of the unstrained configuration. Since the expansion coefficients $U^{(2)}$ and $U^{(3)}$ are related to elastic moduli as summarized in Tab.\ref{table1}, a straightforward fit of 
Eq.(\ref{fit}) has provided the full set of linear moduli and third order elastic constants, while the shear deformation was used to confirm the isotropy of the lattice in the linear approximation.  Each energy-vs-strain curve, shown in Fig.\ref{energy}, has been computed by TB-AS as above described, by increasing the magnitude of $\zeta$ in steps of $0.005$ up to a maximum strain $|\zeta_{max}|=0.055$. Arrows in Fig.\ref{energy} indicate the different nonlinear behavior along the $zz$ and $ac$ directions.  A similar fitting procedure was carried out by computing all the components of the stress tensor for the homogeneous deformations, obtaining no quantitative difference in the calculated moduli.

\begin{table}[b]
\caption{\label{table3} Linear and nonlinear elastic moduli of graphene in units of Nm$^{-1}$ (we remark that the Poisson ratio $\nu$ is dimensionless). }
\begin{tabular}{lccccc}
\hline
 \hline
 \ 		& $ E $ & $ \nu $ &$ D $	& $\langle D_{\vec{n}}\rangle$  & $\langle D^{\mathcal{L}}_{\vec{n}}\rangle$ \\
 \hline
   Present                                    &   312  	                & 	 0.31                  & -                                      &-582.9	    & -1050.9       \\
   Ref.\cite{lee}$ ^{a} $	            &   340$\pm$40 	& 	  -	                       &     -690$\pm$120          &-              & -  	                \\
   Ref.\cite{zhou,arroyo}$ ^{b} $&   235                     &  0.413			       & - 	      & -             & -                                               \\
   Ref.\cite{michel}$ ^{c} $  	    &   384                     &  0.227			        	      & -             & -  & -                                             \\
   Ref.\cite{kudin}$ ^{d} $ 	        &   345                     &  0.149 			      	      & -             & -  & -                                             \\
   Ref.\cite{gui}$ ^{d} $               &    -                          &  0.173  		        	      & -             & -  & -                                             \\
   Ref.\cite{liu}$ ^{d} $                &    350                     &  0.186  		        	      & -             & -  & -                                             \\    
   Ref.\cite{zhou2}$ ^{d} $         &    -                           &  0.32   		           	      & -             & -  & -                                             \\    
   Ref.\cite{sanchez}$ ^{d} $     &    -                           &  0.12-0.19   		   	      & -             & -  & -                                             \\    
\hline
\hline
\end{tabular}

$ ^{a} $ Experimental, $ ^{b} $ Tersoff-Brenner, $ ^{c} $ Empirical force-constant calculations, $ ^{d} $ Ab-initio

\end{table}

The outputs of the fitting procedure are reported in Tab.\ref{table2} where the full set of third order elastic constants of monolayer graphene is shown. We remark that $ C_{111} $ is different than $C_{222} $, i.e a monolayer graphene is isotropic in the linear elasticity approximation, while it is anisotropic when nonlinear features are taken into account. By inserting the elastic constants $ C_{ijk} $ of Tab.\ref{table2} into Eqs.(\ref{lambda}), (\ref{dz}) and (\ref{da}), we also obtained the nonlinear moduli for both the $zz$ and $ac$ directions. 

In Tab.\ref{table3} we report the values of the calculated elastic moduli, together with the available experimental and theoretical data. The present TB-AS value for $E$ is in reasonable agreement with literature \cite{lee,michel,kudin,liu}, while the value of $\nu$ is  larger than most of the ab-initio results \cite{kudin,liu,gui,sanchez} (but for the result in Ref. \cite{zhou2}). While this disagreement is clearly due to the empirical character of the adopted TB model (where, however, no elastic data were inserted in the fitting data base), we remark that the values of $ \langle D_{\vec{n}}\rangle $ and $ \langle D^{\mathcal{L}}_{\vec{n}}\rangle $ predicted by means of Eq.(\ref{dav}) are affected by only 10\% by varying $ \nu $ among the values shown in Tab.\ref{table3}.

\begin{figure}[t]
 \begin{center}
\begin{tabular}{c}
\resizebox{85mm}{!}{\includegraphics{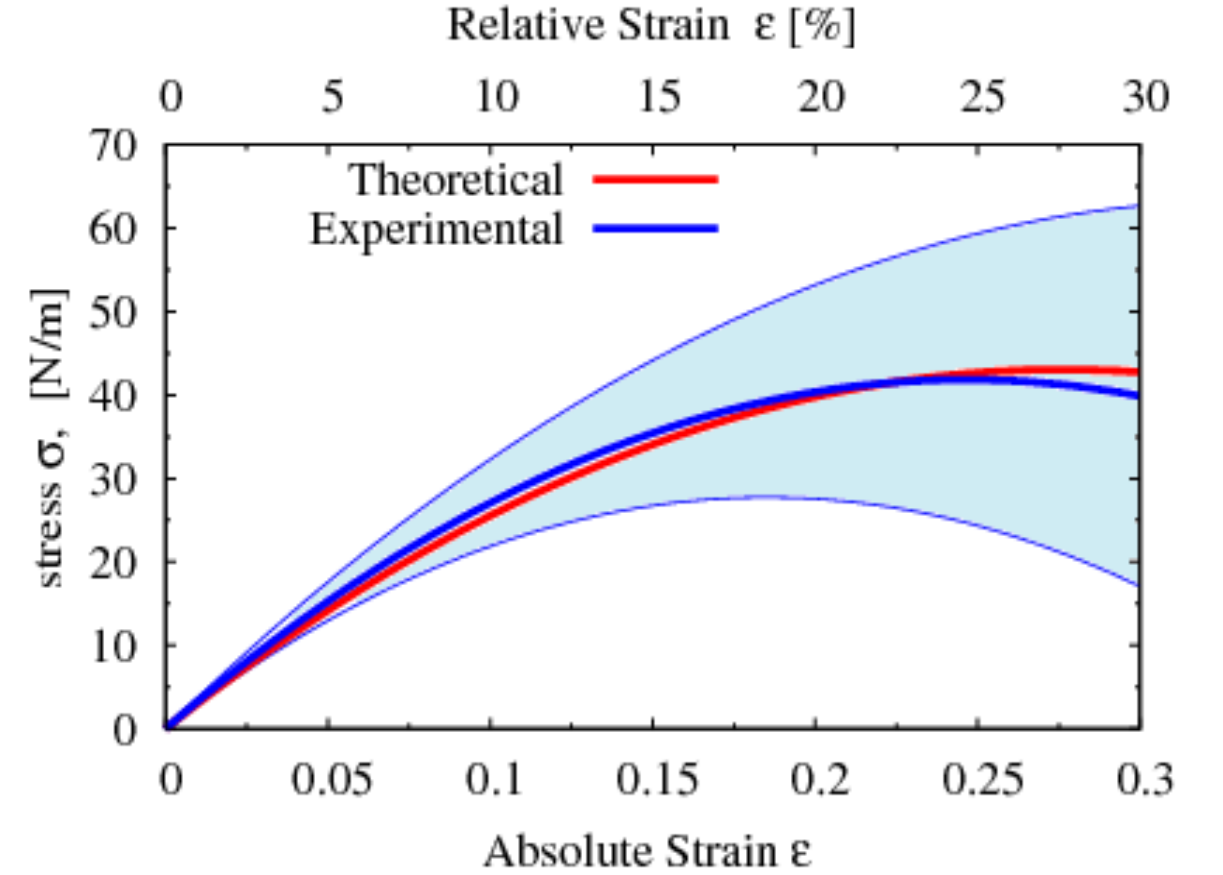}}
\end{tabular}
\caption{\label{curves} (Color online) Theoretical (present work) and experimental (see Ref.\cite{lee}) stress-strain curves, as defined in Eq.(\ref{av}). Shaded area represents the experimental error.}
\end{center}
\end{figure}

Tab.\ref{table3} shows that the predicted $\langle D_{\vec{n}}\rangle$ is much closer to the experimental value $ D $ than its Lagrangian counterpart $\langle D^{\mathcal{L}}_{\vec{n}} \rangle$.  This seems to suggest that, measurements in Ref.\cite{lee} were performed in the physical nonlinearity regime (small strain formalism), rather than in the geometrical nonlinearity one (Lagrangian formalism), as also confirmed by the excellent agreement shown in Fig.\ref{curves} commented below. 
We further observe that the negative sign of all the nonlinear elastic moduli proves that graphene is an hyperelastic softening system (i.e. $D<0$). Therefore, as recently established \cite{gao,Volokh}, the present nonlinear model plays a crucial role in determining the failure behavior of the graphene membrane.

In order to substantiate the above statement, we show in Fig.\ref{curves} the graphene stress-strain curve, as defined in Eq.(\ref{av}). Both the theoretical and experimental curves have been obtained by using the Young modulus and the scalar nonlinear coefficient as reported in Tab.\ref{table3}. We remark that in Fig.\ref{curves} the small strain $\langle D_{\vec{n}}\rangle$ value was used. The agreement between the experimental curve and the theoretical (small strain) one is remarkable. This confirms that likely only physical nonlinearities are at work in the present problem. In addition, by means of Fig.\ref{curves} we can determine the failure stress (maximum of the stress-strain curve) $\sigma_{f}=-E^2/(4\langle D_{\vec{n}}\rangle) $, corresponding to a predicted failure stress as high as  $42.4$  Nm$^{-1}$. This result is in excellent agreement with the experimental value $42 \pm 4$  
Nm$^{-1}$, reported in Ref.\cite{lee}. These values correspond to the failure strength of a two-dimensional system. In order to draw a comparison with bulk materials, we can define an effective three-dimensional failure stress 
$\sigma_{f}^{3D}=\sigma_{f}/d $, where $ d $ can be taken as the interlayer spacing in graphite. By considering $ d=0.335$ nm \cite{jishi}, we obtain $ \sigma_{f}^{3D}\cong130$ GPa. This very high value, exceeding that of most materials (even including other stiff carbon-based systems, e.g. multi-walled nanotubes \cite{peng}), motivates the use of one-atom thick carbon layers as possible reinforcement in advanced composites.

We acknowledge financial support by the project MIUR-PON ''CyberSar''.

\bibliographystyle{apsrev}

\end{document}